# Atomic scale strain engineering on surfaces of two-dimensional materials

N. Sarkar[1*], P.R. Bandaru,[1,3] R.C. Dynes[2,3*]

[1]Department of Mechanical Engineering, [2]Department of Physics,

[3]Program in Materials Science,

University of California San Diego, La Jolla, California 92093-0411, USA

Atomic modulations of two-dimensional materials using scanning tunneling microscope (STM) tip-induced forces modifies their mechanical and electrical properties. In situ topographic and spectroscopic probing through electrical tunneling has been used for straining sheets of graphite, monolayer graphene and $NbSe_2$. The findings also resolve a thirty-five-year-old controversy involving numerous proposed models to explain the source of anomalously high measured atomic amplitudes (of up to 24 Å, expected 0.2Å) from atomic corrugation on graphite surfaces. Our findings attributes the anomaly to surface elastic deformation characteristics of 'soft' 2D monatomic sheets of graphene and graphite in contrast to $NbSe_2$ which is associated with their local bonding configurations. The tip-induced deformations are shown to induce controlled strain on the material surface atomically and it offers a new way for strain engineering. Topographic deformation of formed graphitic Moiré patterns reveals the inter-layer van der Waals (vdW) strength varying across its domains. In-situ tunneling spectroscopy associated with straining of the Moiré domains reveals electronic flat band formation controllably thereby creating a platform for many-body correlations. The paper cautions anomalous observations when probing 2D materials at small gap distances with their strain induced effects and provides guidelines to exploit or avoid this effect.

Such atomic deformation mechanisms are applicable in the emergent fields of twistronics [1] and straintronics [2]. Other conventional methods of using strain as a degree of freedom has brought forth unexpected characteristics in 2D materials *e.g.,* superconductivity [3] and ferromagnetism [4] in bilayer graphene and chalcogenides through hetero-strain[2], hydrostatic pressure, or substrate engineering [5]. Our paper offers another method to induce measurable strain involving the controlled elastic deformation of atomic bonds through STM electrical tunneling induced forces[6-9]. The type and magnitude of the strain in different 2D systems is observed to be a function of the tip-sample interaction and local atomic bonding configuration. The ease of exfoliation of the bulk layered materials and their resultant surface features is related to their structural anisotropy. The exfoliated constituted heterostructures[1] would also be dependent on such issues and need to be investigated for technological implications.

*Electronic address: rdynes@ucsd.edu

We show that scanning tunneling microscopy and spectroscopy can yield significant insights into such strain-based effects. For instance, a large (/small) tip-sample gap as monitored through a small (/large) tunneling current will induce small (/large) deformation of the 2D layer: Fig. 1(a). Quasi-two-dimensional structures with softer surface elastic properties that differ significantly from rigid 3D materials, demonstrates STM's potential to probe various tip induced forces that are not easily accessible due to the conflation of various electronic and mechanical tip-sample interactions [10,11]. These can be deconvoluted through present experiments and analyses of two prototype 2D materials, with the attribute of one- and two-atom based layers, respectively *i.e.,* through graphitic sheets and $NbSe_2$.

Moreover, highly oriented pyrolytic graphite (HoPG) has been extensively used in STM-based work as a calibration standard [12,13] for height determination due to it's known step heights arising from its well recognized atomic flatness and surface cleanliness. However, previous STM measurements have indicated some confusion on it's atomic amplitudes. For instance, while the measured atomic amplitudes on graphite surfaces are expected to be 0.2Å from Local Density of States (LDOS) calculations [14-16], abnormally high atomic amplitudes up to 24Å were reported [6,7,9]. Since the oscillations follow the modeled LDOS based contours[16], earlier electronic models [14-16] based on tip geometry were inadequate to quantitatively account for the anomalously high atomic amplitudes. Prevailing compressible dirt-based model [7] also contradicts our observations. Such anomalies have been attributed to the tip-surface interactions which may induce local elastic deformation [6,8] of graphite surfaces particularly at small gap distances based on an interatomic potential model [17,18] but lacked enough experimental evidence. Hence, the anomalous atomic amplitudes are investigated in this paper across monatomic step heights of graphite (in figure 1b) and $NbSe_2$ (in figure 1c) at different tip-sample gap distances.

Topography across monatomic step heights of graphite (in figure 1d) and $NbSe_2$ (in figure 1e) is shown as a function of tunneling gap distances. To ensure the same magnitude of tunneling gap reduction on both surfaces, the same tip was used over a fixed range of tunneling current or sample bias. An enhancement of atomic amplitudes on graphite and $NbSe_2$ was observed but the step heights remain constant at 3.2 Å and 12.5 Å respectively (corresponding to their c-axis lattice constants[12,19]). Consequently, the step height-based standard in graphite samples is reconfirmed as an accurate height calibration feature in the STM. Additionally in both systems, the spatial frequency of the oscillations (corresponding to a-b plane lattice constants shown in the inset) remain fixed too. Yet there is the issue of the unexpectedly large atomic corrugations on graphitic surfaces - previously attributed, somewhat tentatively [6], to tip-induced surface deformation. The observed surface variation [7] is envisioned by sliding of the basal planes with shear along 2-D layered materials and would be materials specific.

The atomic corrugations can be understood by the individual outward deformation of atoms by tip, as further indicated in Fig. S1 of the *Supplementary Information S1.* Comparison of the topography at two different tip-sample distances shows hills (/valleys) that yields a guide for the indicated placement of the atoms corresponding to outward (/inward) atomic deformations. Closer (/further) gap distances at larger (/smaller) currents corresponding to larger (/smaller) amplitude indicates atomic corrugation but with constant step height. A rationale, from elastic theory [6], for the higher corrugations through tip induced local deformation at an atom suggests outwards (/inwards) deformation related to attractive (/repulsive) tip-sample interactions which would be opposed by a weaker (/stronger) restoring force. This difference was observed [6] in terms of a preference towards larger outward deformation than inward from atomic topography of graphite: see *Supplementary Information S2.*

To glean further insight, the measured atomic amplitudes of graphitic surfaces, $NbSe_2$ and Single layer graphene (SLG) and their average is compared as a function of $I_{tip}$ (at a constant $V_{sample}$): Fig. 2(a), and $V_{sample}$ (at a constant $I_{tip}$): Fig. 2(b). With an increase in the $I_{tip}$ from 10 nA to 90 nA or equivalently a decrease in $V_{sample}$ from 500 mV to 50 mV, an anomalous increase in the atomic amplitudes was observed in graphite from ~ 1.5Å to ~ 3.5Å (as also observed in [6,7,9]) while the step height is relatively constant at around 3.2Å. In contrast, a smaller enhancement was seen in $NbSe_2$ of the atomic amplitudes – in the range of 1.5Å to 2.5Å and SLG – in the range of 2Å to 2.5Å: Fig. 2.
Amplitude enhancements in figure 2 can be best understood when the $I_{tip}$ increment and $V_{sample}$ reduction is seen as consequential reduction in tip-sample gap (shown by arrow in figure 2 plots). A reduction in tip-sample gap results from diminution in tunneling resistance: $R_{gap}$ (defined as the $V_{sample}$ / $I_{tip}$) - also see Fig. S3 in *Supplementary Information S3.* Irrespective of the material, 10 MΩ becomes the onset of tip-induced deformation (as seen in Fig. S3) thus indicating that $R_{gap}$ would need to be greater than 10 MΩ to avoid any tip-induced forces distorting STM topography measurements.

The identification of 10 MΩ is unspecific to materials for onset of deformation from fig. S3, but the extent of deformation is materials specific. It has been suggested [6,9] that the attractive (/compressive) forces from the tip cause an outward (/inward) deformation of the surface which would be manifested through elastic shear along the basal planes: Figs. 3(a) and 3(b), respectively. Consequently, the extent of restoring shear forces is larger in magnitude for a heavier multilayered sheet like $NbSe_2$ compared to a thinner single layer sheet like graphite. Graphite sheets are observed to be one-layer thick separated by ~ 3.2 Å while a $NbSe_2$ surface sheet would be three-atoms (/layers) thick at ~ 12.5 Å. Hence, the extent of deformation and atomic amplitudes would be smaller (/larger) in $NbSe_2$ (/graphite), given similar tip-induced forces. The measured deformations in figure 2 are repeatable and thus within the elastic limit.

A similar tip-surface interaction with the topmost layer of graphite as well as the SLG surface would naively expect to yield comparable deformation. However, the SLG placed on an atomically smooth (ref. Fig. 2) mica substrate, shows much lower atomic amplitude from deformation. This suggests that the anomalous corrugations of graphite surface are not just a surface interaction but involve bonding configurations from the bulk. The vdW bonding between the layers in graphite is distinctly weaker than the stronger coulomb bonding of the SLG to the mica [20]. Indeed, the extent of deformation from figure 2 is seen to be similar for $NbSe_2$ and SLG, suggesting that the SLG-mica adhesion may be comparable in magnitude to vDW strength in $NbSe_2$. Consequently, tip-sample mediated deformation is suggested as a possible methodology for measuring surface-bulk interactions on an atomic scale.

As indicated earlier, the atomic constitution and bonding as well as the structural anisotropy determine the extent of layer deformation and observed atomic amplitudes. We show that such an aspect also manifests in the *tearing* pattern observed after exfoliation. In the case of graphite, a *serrated* step feature with sawtooth waveform-like edges of 10 Å amplitude was seen, *e.g.,* in Fig. 3(e) whereas the $NbSe_2$ step terraces are fractured on the scale of 100 Å: Fig. 3(f) inset. The triangular edges in the graphite follow the *a* and *b* crystal axes - indicated by red and blue arrows, respectively in Fig. 3(e), with respect to the step edge/exfoliation direction (labeled by a white arrow). The exfoliation direction is more closely aligned to *b* thus implying more bond breaking in that direction than along a. Such an exfoliation pattern creates continuous zigzag edges that are applicable for flat band physics (also see *Supplementary Information S4*). Alternatively, in $NbSe_2$*,* there is a propensity for one-sheet (of triatomic-layer) deep pitting at both the micron and nanometer scales.

In surface deformation of 2D material-based sheets, the outstanding features are linked to the structure of the sheet, *i.e.,* number of layers and atomic makeup, as well as the nature of the bonding of the topmost surface layer to the near surface layers. Such behavior is in contrast to what would be nominally expected in a standard constant current STM mode [21], where the tip traces the potential contours over the rigid atomic surface without deforming it. It would then be of interest to investigate the deformation mechanisms for layers slightly out of registry with respect to each other to test the elastic deformation theory and yield more evidence for the model proposed here. Such an aspect has recently found much favor through the study of Moiré patterns, with spatially variable inter-layer bonding strength[22-25]. Twisting two sheets of graphene creates a Moiré pattern with two kinds of domains broadly: (i) AA or BB type stacking, i.e, the A or B atoms are directly on top of each other, or (ii) AB or BA stacking, i.e, the A (/B) atom occupy the hollows of the lower layer hexagons. It is well known that the latter A-B/B-A based Bernal stacking is favored from an energetic stability point of view [22-24] and AA/BB stacking is most unfavorable. However, when twisting one layer with respect to the other, both stacking domains emerge and contribute to the observed Moiré. The pattern periodicity is directly pertinent to the extent of misregistration which is also related to the inter-layer vDW strength. The twistronics implication of such orientability has been manifested in structural, electronic, and physico-chemical attributes [22-24,26-30].

We demonstrate that the tip-induced deformation across the Moiré pattern reveals local vdW bonding strength variability [6,26] as well as interlayer interactions [22-24]. An instance of a pattern formed by a 4.2° twist angle in the upper layers of graphite is shown in Fig. 4(a)**.** AA domains are the hilly regions and AB domains are the valley regions: Fig 4(b). The lines connecting the AA regions are termed domain walls (DWs). Such DWs are of much interest due to specific topological characteristics [26], favorable for soliton propagation [27], *etc.* Scanning the tip in various directions, *e.g.,* AA → AA → AA, along the DW: blue dotted line, and AB → AA→ AB: red dotted line, in Fig. 4(c) and Fig. 4(e) respectively shows that the high (/low) atomic amplitude in the AA (/AB) domains are ascribed to the weak (/strong) vdW-related bonding while the DW bonding is seen to be intermediate to the two modes of stacking. The two topographies are monitored as a function of $I_{tip}$: Fig. 4(c) and 4(e). The deformation amplitudes of the domains as estimated from Figs. 4(c) and 4(e) are indicated in Fig. 5(a) **-** as a function of the $I_{tip}$ and $R_{gap}$. At a reduced tip-sample distance, there is considerable outward deformation*/bulging* out of the AA regions by ~11Å owing to unstable stacking whereas the AB regions are much less deformed, *i.e.,* at ~3Å given its stable stacking configurations [22-24]. The DW response is intermediate, as also previously indicated. The relative deformation of the various moire domains reveals their inter-layer VDW strength. This can be exploited to probe the inter-layer bonding strength of other graphite features. Also, a topographic map of the moiré pattern can be imagined as a map of van der Waals strength.

The differential conductance (*dI/dV*) –corresponding to the local DOS, measured by probing a Moiré pattern, formed at a lower twist angle (of ~ 0.8°) is shown in Fig. 5(b)**.** Such low angle twists have been associated [29] with the occurrence of van Hove singularities (vHs) [24,30] which converge into a single peak, representing a high DOS flat band on deformation: Fig. 5(b) bottom to top inset. A flat band implies minimal energy dispersion. Topography reveals mechanical deformation of moire domains at smaller gap distances which strains the domains. In-situ spectroscopy on the AA domains as a function of deformation reveals the onset of such flat band formation, as deduced from the reduced full width half maxima (FWHM) of the associated conductance peaks, with decreasing $R_{gap}$. Also see

*Supplementary Information S5,* for further discussion on Moiré patterns as well as the interpretation of the measurements. It is then indicated that the deformed top layer has not electronically decoupled from the underlying layer and mimics the deformation indicated in Fig. 3(b) rather than the one in Fig. 4(d).

This paper has demonstrated the interplay between inter-layer coupling as well as the association with the substrate of exemplar one-atom and two-atom constituted 2D material systems, in response to electromechanical forces at the nanoscale. This was achieved through an analysis of the atomic scale corrugations and ascribed to the modulation of physico-chemical bonding attributes. The aspect that band structure tuning would be feasible through the controlled surface deformation would be of importance in the in situ tuning and probing of materials properties.

**Methods:**

## Materials, Exfoliation and Layer synthesis

Highly oriented pyrolytic graphite (HoPG) and niobium diselenide ($NbSe_2$) were freshly cleaved from crystals before loading. The choice of HoPG with roughly micron sized grains (ZYH grade from Advanced Ceramics) favors a dense occurrence of step heights as well as multiple moirés patterns of different twist angles. Single crystals of 2H-$NbSe_2$ millimeter grain sizes were grown by chemical vapor deposition at AT&T Bell labs. Single layer graphene was grown by chemical vapor deposition (CVD) of methane precursor on copper foil and then polymer-assisted wet transfer of graphene was done onto cleaved mica substrates. After dissolving the polymer and thermally evaporating Ti/Au contacts, the graphene/mica stack was baked in an oven at 250$^{o}$ C for a week to dehydrate and degas before loading into the STM.

## Electrical Measurements

All imaging was done using a custom-built tunneling microscope with a RHK controller at room temperature and atmospheric pressure. All topography measurements have been performed in standard constant current mode at ~0.5 Hz scanning frequency using a mechanically snipped Pt/Ir tip. The spectroscopy measurements utilized a lock in modulation of 3mV at 5kHz.

## Calibration and Data analysis

Topographic heights of all images were calibrated using the constant step height measured on graphite and $NbSe_2$ monatomic steps that remains constant irrespective of tunneling conditions. The same tip was used to measure the average atomic heights of graphite, $NbSe_2$ and SLG. Step heights are measured by the difference in mean of the two waveforms on either side of the step. Atomic heights of all materials are averaged over 20 peak-to-peak amplitudes. The conductance peak amplitudes in Fig. 5b has been normalized to show the change in its FWHM (or, flat band tuning) as a function of deformation only.

**References**

1. Soler, J. M., et al. Interatomic forces in scanning tunneling microscopy: giant corrugations of the graphite surface. *Physical review letters* **57**,444 (1986)
2. Mamin, H. Jonathon, et al. Contamination-mediated deformation of graphite by the scanning tunneling microscope. *Physical Review B* **34**,9015 (1986).
3. Pethica, J. B. Comment on ‘‘Interatomic forces in scanning tunneling microscopy: Giant corrugations of the graphite surface’’ *Physical review letters* **57**, 3235 (1986).
4. Koshino, Mikito, and Edward McCann. Multilayer graphenes with mixed stacking structure: Interplay of Bernal and rhombohedral stacking. *Physical Review B* **87**, 045420 (2013).
5. Latychevskaia, Tataiana, et al. Stacking transition in rhombohedral graphite. *Frontiers of Physics* **14**, 1-7 (2019).
6. Bistritzer, R., & MacDonald, A. H. Moiré bands in twisted double-layer graphene. *Proceedings of the National Academy of Sciences* **108**, 12233-12237 (2011).
7. Wu, D., Pan, Y., & Min, T. Twistronics in graphene, from transfer assembly to epitaxy. *Applied Sciences* **10**, 4690 (2020).
8. Miao, F., Liang, SJ. & Cheng, B. Straintronics with van der Waals materials. *npj Quantum Mater* **6**, 59 (2021).
9. Cao, Y., Fatemi, V., Fang, S. et al. Unconventional superconductivity in magic-angle graphene superlattices. *Nature* **556,** 43–50 (2018).
10. Sharpe, A. L. et al. Emergent ferromagnetism near three-quarters filling in twisted bilayer graphene. *Science* **365**, 605–608 (2019).
11. Yang, S., Chen, Y., & Jiang, C.. Strain engineering of two-dimensional materials: Methods, properties, and applications. *InfoMat* **3**, 397-420 (2021).
12. Binnig, Gerd, et al. Energy-dependent state-density corrugation of a graphite surface as seen by scanning tunneling microscopy. *Europhysics Letters* **1**, 31 (1986).
13. Dürig, U., J. K. Gimzewski, and D. W. Pohl. Experimental observation of forces acting during scanning tunneling microscopy. *Physical review letters* **57**, 2403 (1986).
14. Mate, C. Mathew, et al.Direct measurement of forces during scanning tunneling microscope imaging of graphite. *Surface science* **208**, 473-486 (1989).

15. Meyer, Jannik C., et al. On the roughness of single-and bi-layer graphene membranes. *Solid State Communications* **143**, 101-109 (2007).
16. Batra, Inder P., et al. A study of graphite surface with STM and electronic structure calculations. *Surface Science* **181**,126-138 (1987).
17. Wong, H. S., Durkan, C., & Chandrasekhar, N. Tailoring the local interaction between graphene layers in graphite at the atomic scale and above using scanning tunneling microscopy. *ACS nano*. **3**, 3455-3462 (2009).
18. Tersoff, J. Anomalous corrugations in scanning tunneling microscopy: imaging of individual states. *Phys. Rev. Lett.* **57**, 440 (1986).
19. Tersoff, Jerry, and Donald R. Hamann. Theory of the scanning tunneling microscope. *Physical Review B* **31**, 805 (1985).
20. Tekman, E., and S. Ciraci. Atomic theory of scanning tunneling microscopy. *Physical Review B* **40**, 10286 (1989).
21. Olsen, Martin, Magnus Hummelgård, and Håkan Olin. Surface modifications by field induced diffusion. *PLoS One* **7**, e30106 (2012).
22. Marezio, M., et al. The crystal structure of NbSe2 at 15 K. *Journal of Solid State Chemistry* **4**, 425-429 (1972).
23. Shim, Jihye, et al. Water-gated charge doping of graphene induced by mica substrates. *Nano letters* **12**, 648-654 (2012).
24. Binnig, Gerd, and Heinrich Rohrer. Scanning tunneling microscopy. *Surface science* **126**, 236-2441 (1983).
25. Gong, Lei, et al. Reversible loss of bernal stacking during the deformation of few-layer graphene in nanocomposites. *Acs Nano* **7**, 7287-7294 (2013).
26. Huang, Shengqiang, et al. Topologically protected helical states in minimally twisted bilayer graphene. *Physical review letters* **121**, 037702 (2018).
27. Alden, Jonathan S., et al. Strain solitons and topological defects in bilayer graphene. *Proceedings of the National Academy of Sciences* **110**, 11256-11260 (2013).
28. McGilly, Leo J., et al. Visualization of moiré superlattices. *Nature Nanotechnology* **15**, 580-584 (2020).
29. Liu, Xiaomeng, et al. Spectroscopy of a tunable moiré system with a correlated and topological flat band. *Nature communications* **12**, 1-7(2021).
30. Kerelsky, Alexander, et al. Maximized electron interactions at the magic angle in twisted bilayer graphene. *Nature* **572**, 95-100 (2019).

## Acknowledgements:

This work was supported by AFOSR Grant (FA9550-15-1-0218) and Army Research Office (AROW911NF-21-1-0041). The authors wish to thank Michael Rezin for the technical assistance; Prof. Shane Cybart, Uday Sravan Goteti and Hidenori Yamada for useful discussions.

## Author contributions:

N. Sarkar did the experimental work and along with P.R. Bandaru and R.C.Dynes, wrote the paper. All analysis and discussion were under the supervision of R.C. Dynes and P.R. Bandaru.

## Data availability:

The experimental data and its analysis in the paper and/or in the supplementary information is sufficient to support our conclusions. Additional data can be made available on request.

## Competing interests:

The authors declare no competing interests.

# Figures

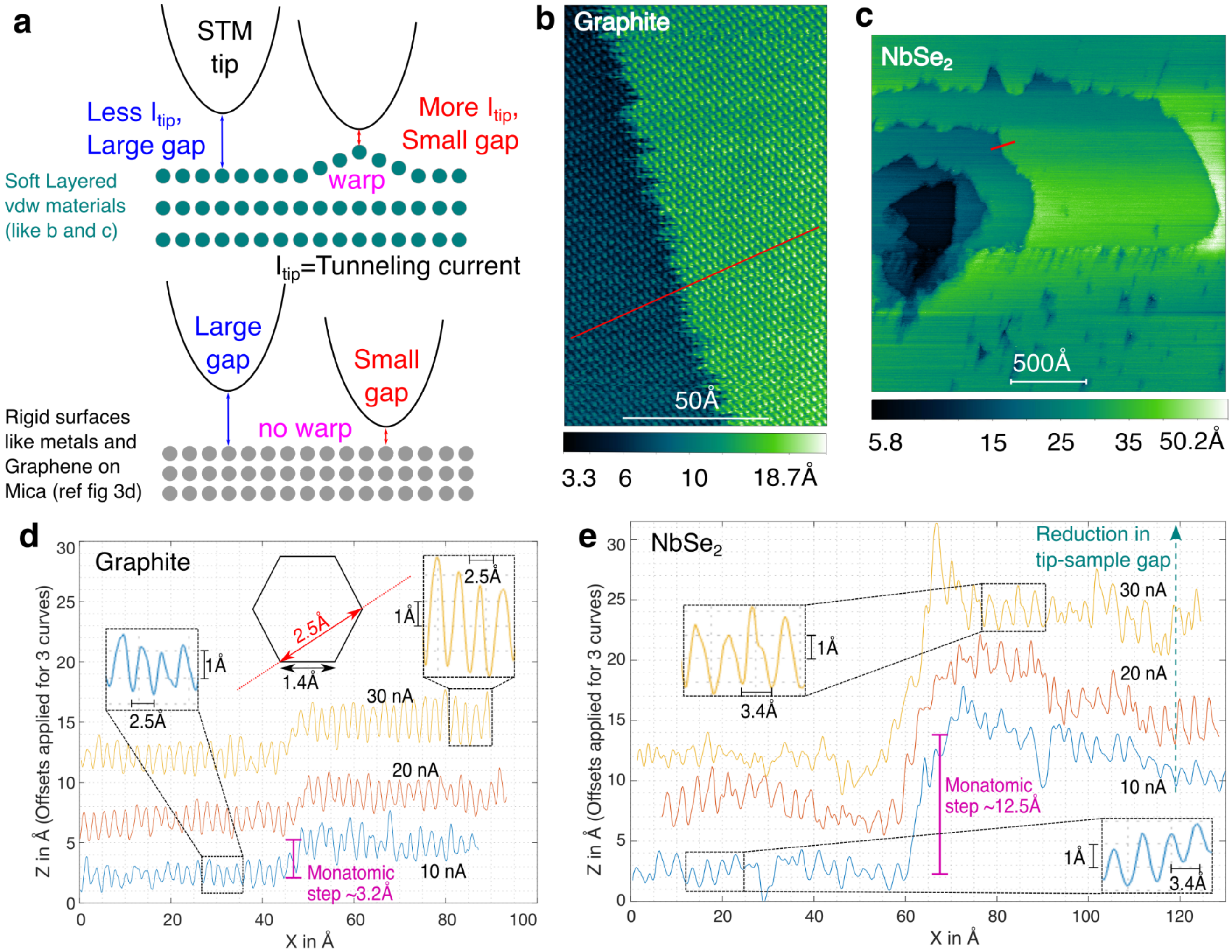

Figure1: **Tip-induced atomic deformation of graphite and $NbSe_2$ surfaces**. **(a)** The expected response of *rigid* vs. *soft* materials to tip induced deformations at large (/small) tip-sample gaps monitored by higher (/small) tunneling currents in scanning tunneling microscopy (STM). The topographic images along steps on **(b)** graphite surface (100Å x 145Å) and **(c)** $NbSe_2$ surface (2300Å x 2300Å) - at 30nA tunneling current ($I_{tip}$) and 100mV sample bias ($V_{sample}$). The color bar at the bottom indicates the measured height in angstroms (Å). The atomic corrugations across the step, as a function of $I_{tip}$ topography along the red line in **(b)** and **(c)** for **(d)** graphite, and **(e)** $NbSe_2$.The oscillations correspond to corrugations from scanning across individual atoms. The insets show in a magnified view the increase in average oscillation amplitude from 1.5Å to 2.7Å for graphite in **(d)** and from 1.7Å to 2.4Å for $NbSe_2$ in **(e).** The spatial frequency on either side of the step remains constant at 2.5Å for graphite - corresponding to the shown atomic spacing in the a-b plane in the middle inset in **(d)** and at 3.4Å for $NbSe_2$ in **(e)**. The oscillation amplitudes and step heights calculated from (d) and (e) are plotted in figure 2 as a function of $I_{tip}$ and $V_{sample}$.

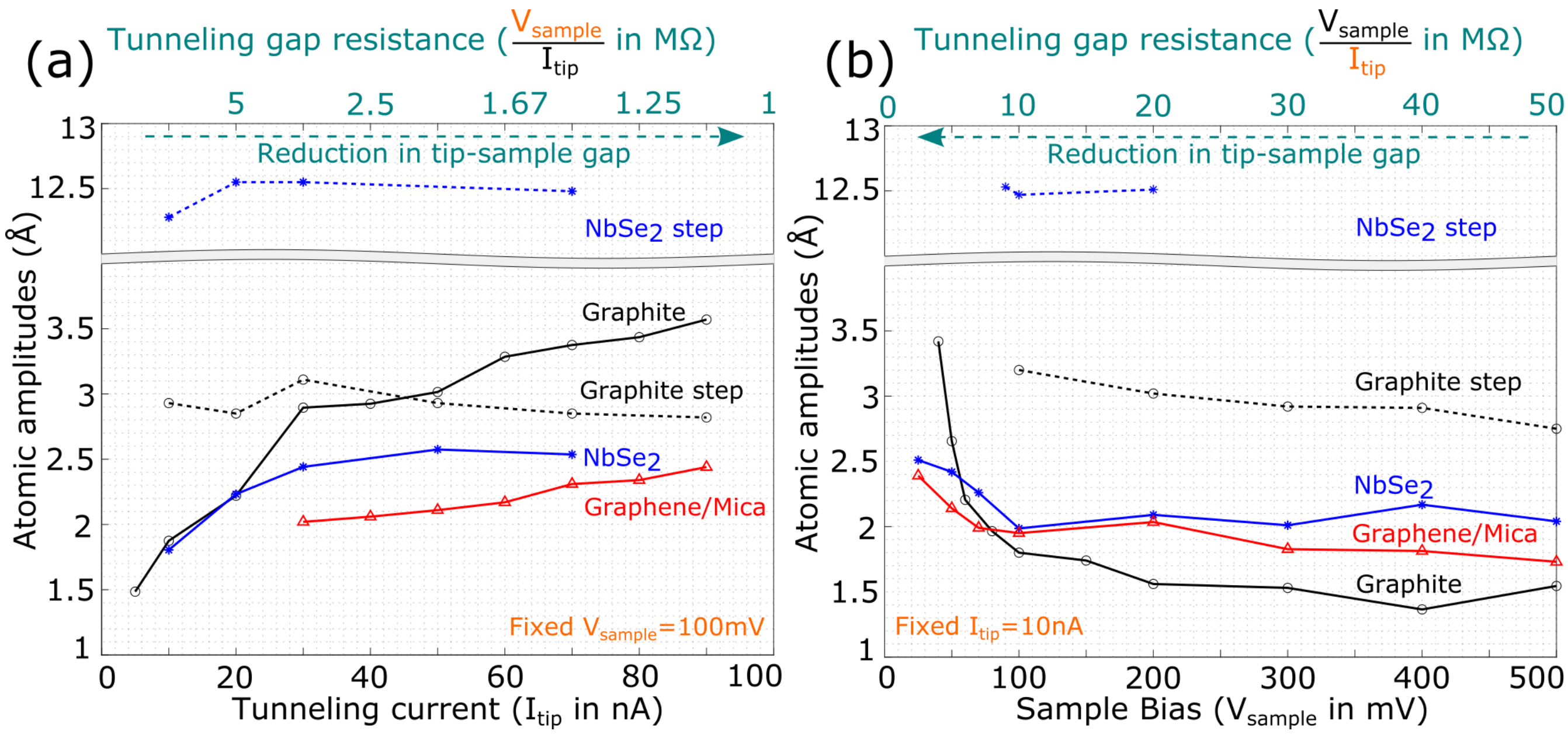

Figure 2. **The atomic corrugation amplitude as a function of scanning tip – surface spacing**, An increase in the **(a)** tunneling current: $I_{tip}$, or equivalently a decrease in **(b)** voltage applied to the sample 2D material: $V_{sample}$, is related to a reduced spacing, and yields higher amplitude of atomic oscillation. A larger increase is found on the graphite surface compared to $NbSe_2$ and monolayer graphene. The tunneling gap resistance (defined through the ratio of $V_{sample}/I_{tip}$) – on the top horizontal axis, reduces with a decreased tip-surface spacing.

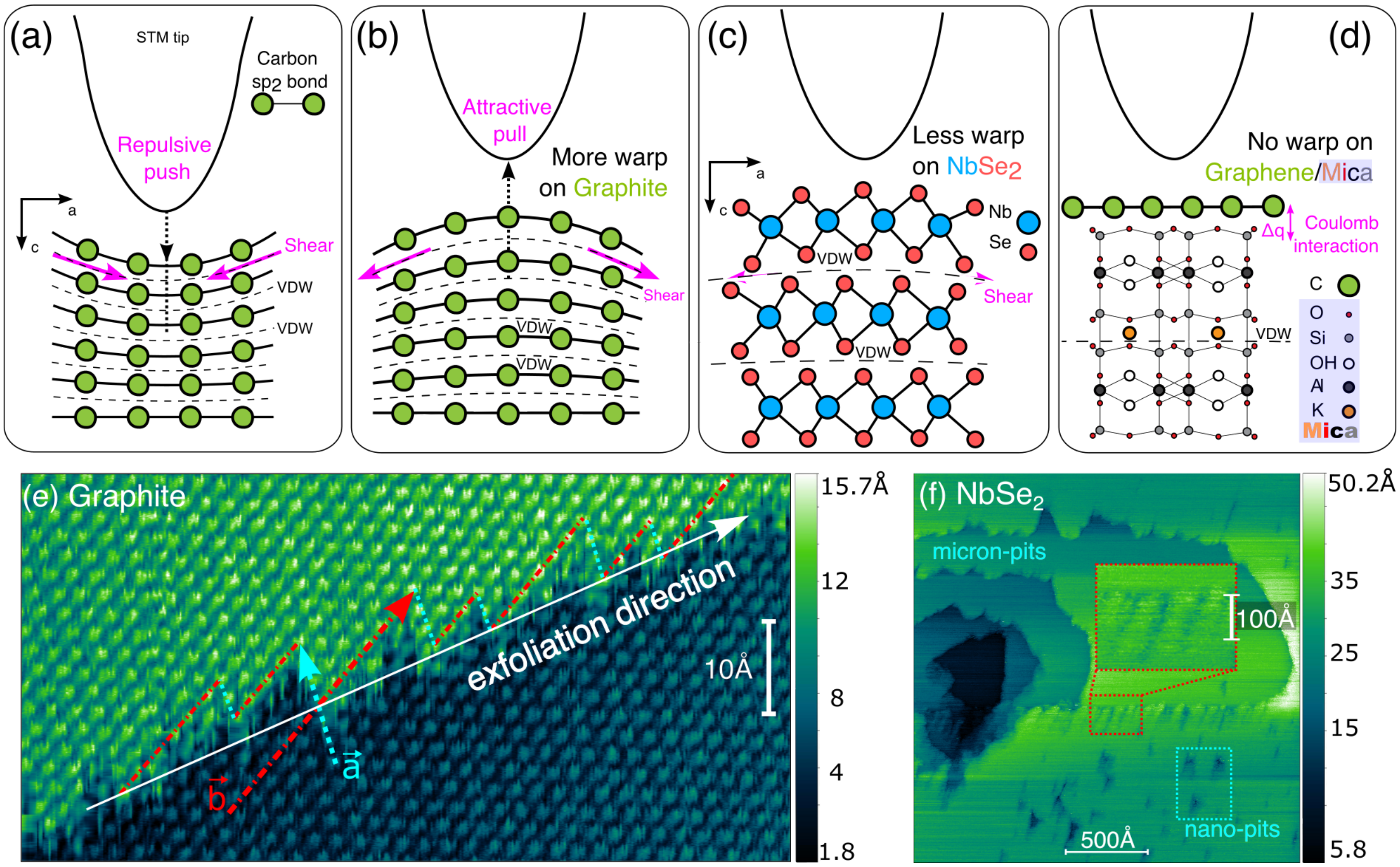

Figure 3: **Structural anisotropy explains the elastic deformation amount and exfoliation patterns**. Tip-induced, elastic, local **(a)** compression, and **(b)** expansion of graphite surface corresponding to repulsive and attractive interactions, respectively. The related reaction from the underlying near-surface layers is depicted as a shear between the basal planes. deformations are opposed by shear forces (arrows) along basal planes. In contrast to the representative one-layer surface of graphite the **(c)** $NbSe_2$ sheet is composed from three atomic sheets corresponding to Se-Nb-Se. The stronger bonding in the case of the $NbSe_2$ sheets is comparable to that observed for **(d)** the single layer graphene (SLG) on a mica substrate, as deduced from the similar atomic corrugation amplitudes, *cf.,* **Fig. 2** in both cases, and is manifested in a much smaller warp compared to the case of graphite – where there is substantial bowing and enhanced amplitude. Exfoliation of 2D sheets results in bond breaking and extensive localized plastic deformation and is manifested as a **(e)** serrated pattern in the atomically resolved topography of graphite, near a step edge. The tearing occurs along the ***a*** and ***b*** crystal axes and creates a sawtooth-like pattern. Alternately, a series of monatomic step heights are observed on exfoliated **(f)** $NbSe_2$ surfaces, with V-shaped pitting at the micron- and nanometer scales.

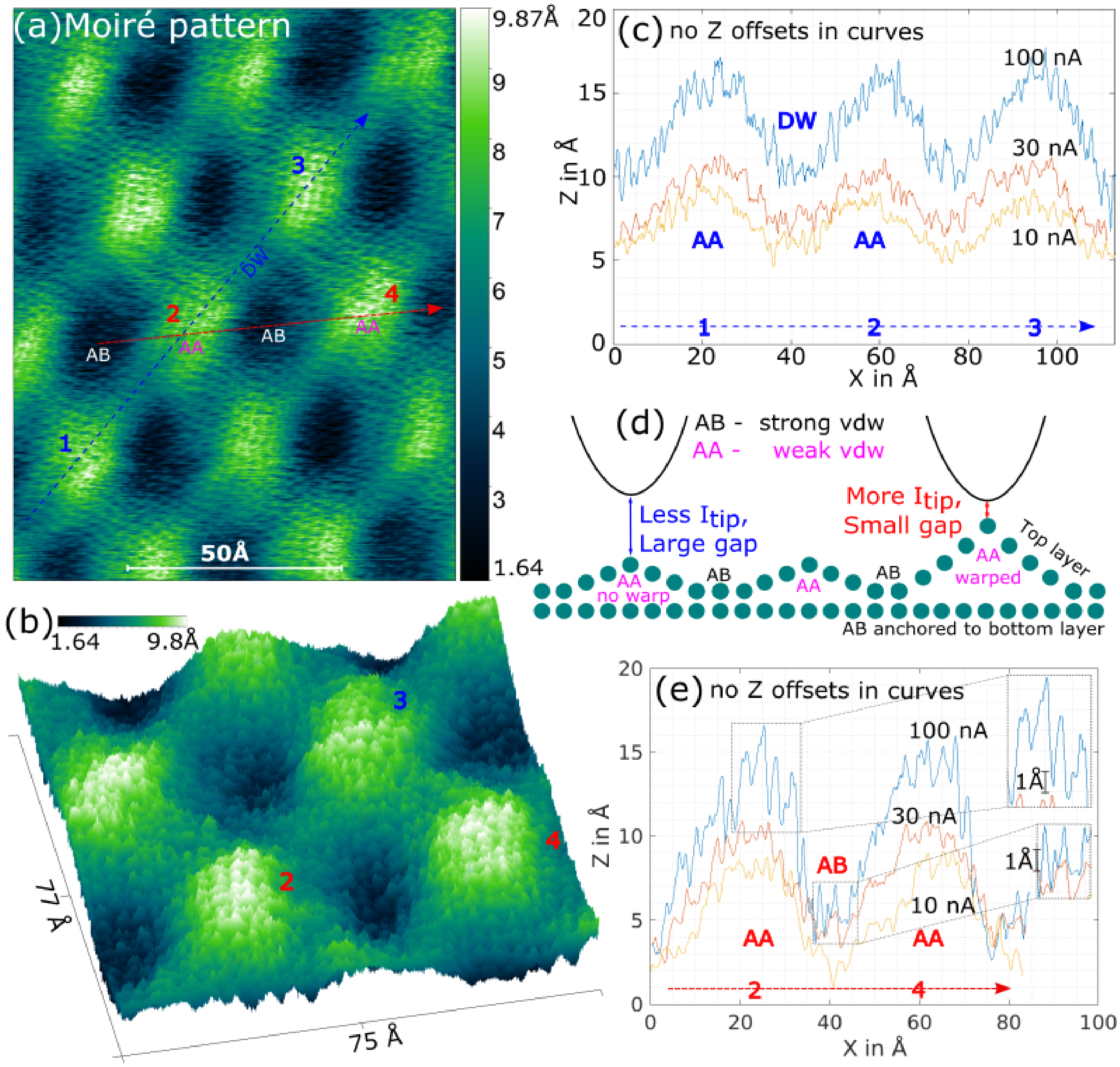

Figure 4: **Probing interlayer interactions and bonding variability in Moiré pattern domains. (a)** Two- and **(b)** Three-dimensional atomically resolved topography of Moiré pattern in graphite with a 4.2° twist angle between the top and the bottom layer. A pattern of *bright* domains (termed 'AA') and *dark* domains (termed 'AB') joined by domain walls (DW) are noted and correspond to the atomic arrangements related to the stacking configuration of the individual graphene layers. Two distinct scan directions (AA → AA → AA, along the DW: blue dotted line, and AB → AA → AB: red dotted line) are shown, with the corresponding atomic amplitude variations at differing $I_{tip}$ in **(c).** A high (/low) atomic amplitude is observed in the AA (/AB) region, due to the weak (/strong) vDW-related bonding. **(d)** The contouring of the surface in the Moiré pattern is reflected in smaller (/larger) tunneling currents prevalent at larger (/smaller) tip-sample distance. **(e)** A scan along the AA → AB → AA direction, indicating that the AA (/AB) regions are deformed outward (/anchored) to the bottom layer. The *inset* indicates that the AA domains have higher individual atomic amplitudes compared to the AB case The scans in **(c)** and **(e)** are not offset and consequently indicate surfaces being pulled out of plane. The DW was indicated to be of bonding strength intermediate to AA and AB regions.

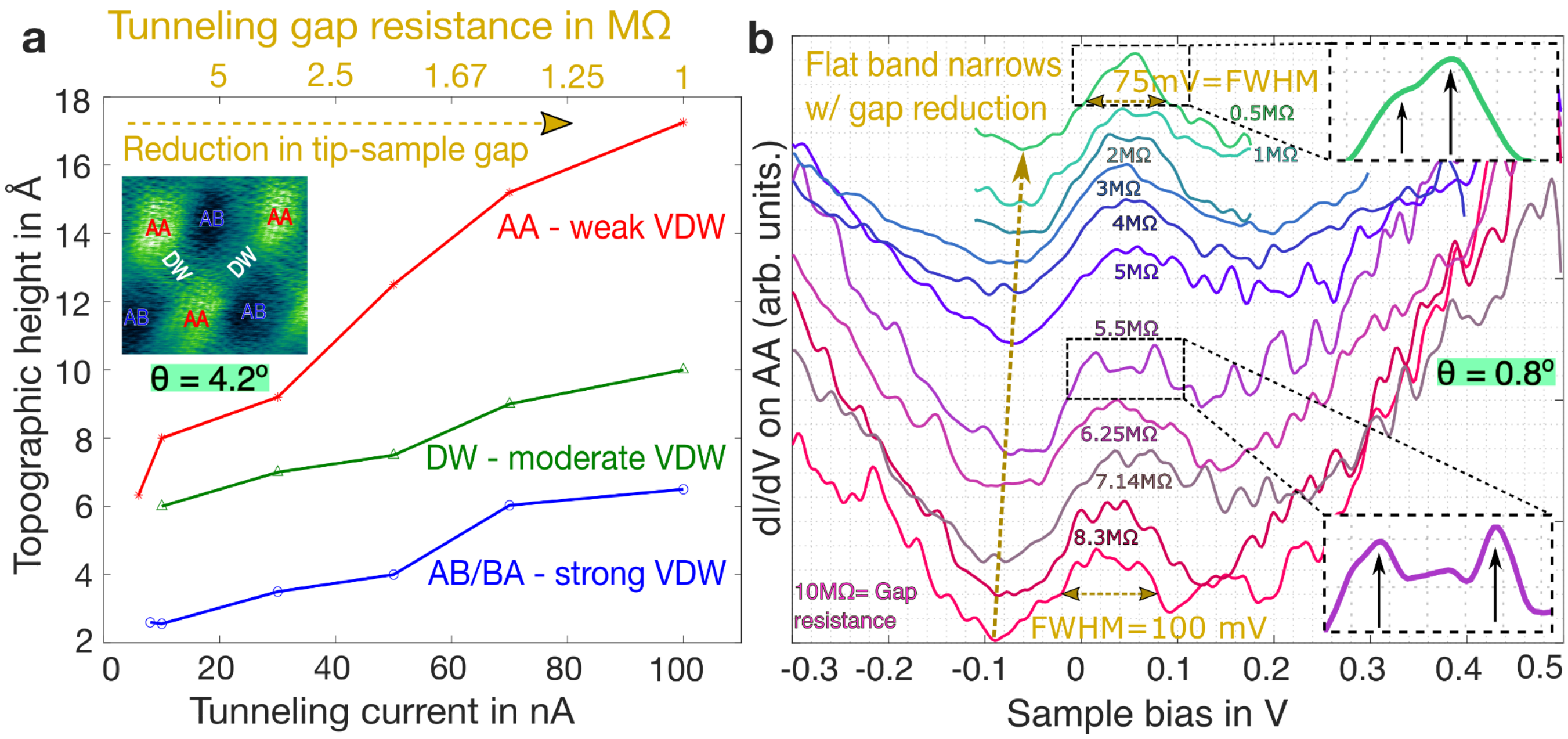

Figure 5: **Topographic and electronic response of Moiré related domains to tip-induced deformations.** **(a)** The measured height of the domains (AA and AB types corresponding to the graphite layer stacking) and the domain walls (DW) increases as the tip-surface gap reduces through the variation of the tunneling current. A larger (/smaller) extent of deformation is seen for the AA (/AB) domains with an intermediate extent for the DWs. **(b)** Conductance (dI/dV) spectrometry on the AA domains of small twist angle (~ 0.8°) indicates the *onset* of flat band attributes, as deduced from the reduction of the peak full width half maxima (FWHM). Here, a reduced tip-sample gap implies greater deformation of the topmost layer and reduced independence of the layer from the bulk.